\documentclass[jcp,twocolumn,showpacs,superscriptaddress,groupedaddress]{revtex4}  
\usepackage{graphicx}  
\usepackage{dcolumn}   
\usepackage{bm}        
\usepackage{amssymb}   
\usepackage{color}

\begin{document}
\title{Kinetic stability and energetics of simulated glasses created by constant pressure cooling}
\author{Hannah Staley}
\affiliation{Department of Physics, Colorado State University, Fort Collins, Colorado 80523, USA}
\author{Elijah Flenner}
\author{Grzegorz Szamel}
\affiliation{Department of Chemistry, Colorado State University, Fort Collins, Colorado 80523, USA}
\date{\today}
\begin{abstract}
We use computer simulations to study the cooling rate dependence of the 
stability and energetics of model glasses created at constant pressure conditions
and compare the results with glasses formed at constant volume conditions. 
To examine the stability, we determine the time it takes for a glass cooled 
and reheated at constant pressure to transform back into a liquid, $t_{\mathrm{trans}}$,
and calculate the stability ratio $S = t_{\mathrm{trans}}/\tau_\alpha$, where 
$\tau_\alpha$ is the equilibrium relaxation time of the liquid.  
We find that, for slow enough cooling rates, cooling and reheating at 
constant pressure results in a larger stability ratio $S$ than for cooling and reheating
at constant volume. 
We also compare the energetics of glasses obtained by cooling while maintaining constant pressure 
with those of glasses created by cooling from the same state point while maintaining constant volume.
We find that cooling  at constant pressure results in glasses with lower
average potential energy and average inherent structure energy. 
We note that  in model simulations of the vapor deposition process glasses are created under constant pressure conditions,
and thus they should be compared to glasses obtained by constant pressure cooling.
\end{abstract}

\maketitle

\section{Introduction}

Vapor deposition of molecules onto a substrate held around 85\% of their glass transition temperature
is used to create glasses whose kinetic stability is much larger than glasses 
created by cooling at a constant rate \cite{E_S}.  
To study the stability of the vapor deposited 
glasses, 
Swallen \textit{et al.} \cite{E_S} compared the heat capacity of highly stable glasses and of 
glasses created by cooling at a constant rate, i.e.\ ordinary glasses, 
while heating these glasses at a constant rate.
From the peak in the heat capacity they determined the onset temperature for melting, and found 
that the onset temperature for the vapor deposited glasses was much higher than for 
the ordinary glasses. 
More recently, a different procedure was used.  
Sep\'{u}lveda \textit{et al.} \cite{Sep} quickly heated vapor deposited glasses to 
a liquid temperature and then held them at that higher temperature. 
They defined the transformation time $t_{\mathrm{trans}}$ as the time after which the response becomes liquid-like, 
and defined the stability ratio $S = t_{\mathrm{trans}}/\tau_\alpha$ where 
$\tau_\alpha$ is the relaxation time of the liquid.  This procedure provided a more quantitative way to characterize 
a glass's stability. 

The discovery of highly stable glasses created by vapor deposition has prompted researchers to devise various simulational 
protocols to create highly stable simulated glasses, study their stability, and examine the
characteristics of a system that would make it a more stable glass. Jack \textit{et al.} \cite{JHGC} found
that more stable glasses would have lower average inherent structure energies than ordinary glasses by  
using the $s$-ensemble to bias inactive states that
were more kinetically stable than other states at the same temperatures. 
A  recent simulational study by Helfferich \textit{et al.}\ \cite{Helfferich2016} demonstrated that the 
average inherent structure energy was a good indicator of the mobility of particles in vapor deposited
and aged simulated glassy films, which suggests that the inherent structure energy may be 
a good indicator of the stability of these films. 

To examine the creation of vapor deposited glasses, L\'eonard and Harrowell \cite{LH} used a 
three-spin facilitated Ising model to model vapor deposition. 
They found that to match the stability of the glasses created by simulated vapor deposition at the
slowest deposition rate, constant cooling rate simulations starting from a bulk system would take $10^6$ times longer. 
Hocky \textit{et al.} \cite{HBR} created stable
two-dimensional glasses using random pinning. They reheated and cooled their pinned glasses at a
constant rate and demonstrated that random pinning created glasses lower in the potential energy landscape.
These simulations used various protocols to examine properties of systems with increased stability, but they were
not designed to examine properties of ordinary simulated glasses. 

To examine the increased stability of vapor deposited glasses versus ordinary glasses, 
in a series of simulations de Pablo, Ediger and collaborators 
examined the stability of glasses created by a protocol based on vapor deposition and 
compared these glasses to simulated glasses created by cooling at a constant rate 
\cite{Helfferich2016,SdP,E_NM,E_JCP,Lin2014,Lyubimov2015}. 
It was found that the vapor deposited glass films were more stable than glass films 
created by cooling at a constant rate. In early studies the vapor deposited glasses were compared to glasses cooled
at a constant rate and at a constant density that was higher than the density of the vapor deposited glasses \cite{E_NM}.
It was later found that composition effects resulted in an over estimation of the stability of the 
vapor deposited glass \cite{E_JCP}. However, it was clear that the vapor deposited glasses were indeed 
more stable than glasses prepared by cooling at a constant rate, and that the inherent structure energy gave
insight into the stability of the glass \cite{Helfferich2016,E_JCP,cool}, 
but one had to be careful in comparing the stability of glasses prepared 
by different methods. Specifically, to examine the stability of glasses created through a vapor deposition 
algorithm it was determined that a suitable procedure is to reheat and cool the vapor deposited film
\cite{Helfferich2016,E_JCP,Lin2014,Lyubimov2015}. However, it is unclear what the effects of the
free surface and the substrate are, and thus it is also informative to examine simulated glasses
prepared by cooling bulk liquids (i.e.\ simulated glasses prepared using periodic boundary conditions
to approximate an infinite system) at a constant rate. 

Since it has been found that simulated vapor deposited glasses are at zero pressure \cite{E_JCP},
the stability of vapor deposited glasses should be compared to 
bulk glasses prepared at constant pressure $P=0$. Lyubimov \textit{et al.}\ performed a
brief study to compare glasses cooled at constant pressure $P=0$ to vapor 
deposited glasses, and found that the average energy was similar to that of
the glass films if they were both cooled at the same rate. However, 
the films had a lower inherent structure energies than the bulk glasses, but
the density of the films increased during the energy minimization procedure and
the density of the bulk glasses did not change. 
Lyubimov \textit{et al.}\ did not perform a detailed comparison of the stability of the
bulk glasses cooled at a constant $P=0$ to the vapor deposited glasses. 

Due to the increased interest in characterizing the stability of simulated glasses formed by different means and
what constitutes a stable glass formed through simulation, we performed a detailed study 
of the stability of a model 
glass forming system created by cooling at constant rate and at constant density \cite{cool}. 
As in previous simulations \cite{JHGC,Helfferich2016,SdP,E_NM,E_JCP,Lin2014,Lyubimov2015}, we found 
that the average energy and the inherent structure energy were lower for more stable glasses \cite{cool}. Furthermore,
we established methods to examine the properties of glasses created in simulations. To asses the stability 
of the glass we used a procedure modeled after the 
experiments of Sep\'{u}lveda \textit{et al.}\ \cite{Sep} and the simulations of Hocky \textit{et al.}\ \cite{HBR}.
To this end we quickly heated the glass to a supercooled liquid temperature and held it at the constant temperature,
and waited until the glass transformed back into a liquid. Note that all the simulations 
in our previous work, Ref. \cite{cool}, were performed at constant density. We then defined a stability ratio
$S = t_{\mathrm{trans}}/\tau_\alpha$ to characterize the stability of the glass. While we found that 
the largest stability ratio that we could achieve, $S = 65.6$,
was much smaller than those for the most stable glasses prepared in simulations, $S \approx 400$,
and for vapor deposited glasses prepared in the lab, $S \approx 3000$, it is unclear how the procedure of 
creating and melting the glass would change the stability ratio. 

Here we examine the stability of glasses created by cooling at a constant rate under constant pressure conditions
instead of constant volume conditions. To determine the stability we monitor the system's relaxation after
a sudden reheating at constant pressure conditions. 
To make a quantitative comparison with previous work, we investigate the same standard model glass-former 
as in Ref.~\cite{cool}, we start from the same state point as in our previous study, and we choose a pressure 
such that the average volume in equilibrium at a liquid state point where we began the cooling is the same as in 
our previous constant volume simulations. 
We find that cooling at constant pressure creates 
glasses that are lower in the potential energy landscape 
and we find larger stability ratios than in the previous study performed at constant volume.
 
 The paper is organized as follows. In Section \ref{sec:simulations} we describe the 
 simulations, the averaging procedure for our out of equilibrium simulations, 
 and checks to make sure that the system did not crystallize. 
 Then in Section \ref{sec:stability} we compare the kinetic stability of
 simulated glasses created by cooling at a constant rate at constant
 volume and constant pressure and in Section \ref{sec:energetics} we discuss the energetic properties of simulated glasses. We summarize the work and 
 draw conclusions in Section \ref{sec:conclusions}.

\section{Simulations}
\label{sec:simulations}
We simulated the 80:20 binary Lennard-Jones mixture introduced by Kob and Andersen (KA) \cite{kob1994,ka,kob1995}. 
The interaction potential
is 
$V_{\alpha\beta} = 4 \epsilon_{\alpha\beta} \left[ \left( \frac{\sigma_{\alpha\beta}}{r}\right)^{12} 
- \left( \frac{\sigma_{\alpha\beta}}{r} \right)^{6} \right]$, with parameters: 
$\epsilon_{AB} = 1.5\epsilon_{AA}$, $\epsilon_{BB} = 0.5\epsilon_{AA}$, 
$\sigma_{AB} = 0.8\sigma_{AA}$, and $\sigma_{BB} = 0.88\sigma_{AA}$. 
The masses of the species are equal, and type A particles are the majority species. 
We present results in reduced units with $\sigma_{AA}$ being the unit for length, $\epsilon_{AA}/k_B$ 
the unit for temperature, 
and $\sqrt{m_A \sigma_{AA}^2 /\epsilon_{AA}}$ the unit for time. 
We simulated $N = 8000$ particles at a constant pressure of $P = 3.958$, 
which is the average pressure of an 
equilibrium system at a number density of $\rho = N/V = 1.2040$ (a box length of 18.8) and a temperature of 0.5. 
We ran NPT simulations with a Nos\'{e}-Hoover thermostat and barostat using LAMMPS (Large-scale Atomic/Molecular Massively 
Parallel Simulator) \cite{L_o,L_a,L_g} and HOOMD (Highly Optimized Object-Oriented Molecular Dynamics)-blue \cite{h_o,h_a}. 
We used a time-step of size 0.002, a thermostat time constant of 0.2 and a barostat time constant of 2.0. Most simulations 
were run on an NVIDIA Tesla K20c GPU (graphics processing unit).

Our systems were out-of-equilibrium, and thus we could not average over time origins. In Subsections \ref{sec:cool} and 
\ref{sec:heat} we discuss the simulations and our averaging procedures. In Subsection \ref{sec:crystal}, we show how we 
checked that our glasses did not form crystals.

\subsection{Cooling}
\label{sec:cool}

We studied glasses prepared by cooling at rate of $\dot{T} = \Delta T/\Delta t$ of $3.33 \times 10^{-n}$ where $n=4$, 5, 6, 7, and 8. We created independent equilibrium configurations at the supercooled temperature of 0.5. We then cooled these independent configurations from $T=0.5$ to $T=0.3$. For all cooling rates except for the slowest cooling rate, $\dot{T} = 3.33\times10^{-8}$, we cooled 80 independent configurations. We cooled 4 independent configurations at the slowest cooling rate due to time constraints. 

From our cooling trajectories we calculated the average potential energy $\left< U \right>$, 
the average inherent structure energy $\left < E_{IS} \right>$, the average density $\rho$, 
the order parameter $Q_6$, and the partial radial distribution functions $g_{\alpha \beta}(r)$ at $T = 0.3$. 
For each trajectory, we averaged several configurations around $T = 0.3$ to 
obtain our non-equilibrium averages for a single run. 
We then averaged the values from different trajectories. 
We used the FIRE algorithm \cite{fi} implemented in HOOMD-blue to quench the 
$T = 0.3$ configurations to their inherent structures. 

\subsection{Heating trajectories}
\label{sec:heat}

We heated the configurations obtained by cooling at a constant rate from $T = 0.3$ to $T = 0.5$ over a time of $t = 10$, a 
small fraction of the total heating trajectory (the heating was done at a constant rate). 
We then continued running while maintaining the temperature at $T = 0.5$ for 
at least as long as it took for the systems to return to a liquid state. We refer to the ramping up of temperature to $T=0.5$ 
and the subsequent run at $T = 0.5$ as a heating trajectory. 
We note that ramping up the temperature over a time of 10 was necessary, 
because an instantaneous increase in temperature resulted in large oscillations of the potential energy due to the thermostat. 

For the cooling rates $3.33 \times 10^{-n}$, where $n = 4$ to 7, we ran 80 heating trajectories from the 
configurations of the cooling runs. For each of the 4 cooling runs at $3.33 \times 10^{-8}$, we ran 
20 different heating trajectories with different initial random velocities. Thus, we also had 
80 heating trajectories at this slowest cooling rate. 
For each cooling rate the results are averages over the 80 heating trajectories.
All the heating trajectories were obtained by running constant pressure simulations. 

\subsection{Checks for crystallization}
\label{sec:crystal}

We checked that our system had not crystallized by examining the spherical harmonic order parameter $Q_6$ and the 
partial pair 
distribution functions $g_{\alpha\beta}(r)$. 

We used the definition of $Q_6$ from Refs.~\cite{E_NM,cool}. First,
we define the complex number
\begin{equation}
\label{eq:qlm}
q_{lm} (i) = \frac{1}{N_b (i)} \sum_{j = 1}^{N_b (i)} Y_{lm} (\mathbf{r}_{ij}),
\end{equation}
for particle $i$, where $N_b (i)$ is the number of neighbors of particle $i$, 
$Y_{lm} (\mathbf{r}_{ij})$ are the spherical harmonics, 
$\mathbf{r}_{ij} = \mathbf{r}_j - \mathbf{r}_{i}$, and $\mathbf{r}_i$ is the position of particle $i$. 
We define the nearest neighbors of particle $i$ as particles within a distance of 1.8 from particle $i$. 
We calculated the local order parameter,
\begin{equation}
\label{eq:ql}
\bar{q}_{l} (i) = \sqrt{\frac{4 \pi}{2l + 1} \sum_{m = -l}^{l} \left| \bar{q}_{lm} (i) \right|^2},
\end{equation}
where
\begin{equation}
\label{eq:qbarlm}
\bar{q}_{lm} (i) = \frac{1}{N_b (i) + 1} \sum_{k = 1}^{N_b (i) + 1} q_{lm} (k).
\end{equation}
In eq.~\ref{eq:qbarlm}, the sum is over the nearest neighbors of particle $i$ as well as particle $i$ itself. 
We define the order parameter $Q_6$ as
\begin{equation}
\label{eq:q6}
Q_{6}  = \frac{1}{N} \sum_{i} \bar{q}_{6} (i).
\end{equation}

Table~\ref{table1} gives the values of $Q_6$ at the different cooling rates for our constant pressure simulations and the 
constant volume simulations of Ref.~\cite{cool}. The values of $Q_6$ are small for both 
the constant pressure and constant volume 
simulations, suggesting that the system did not crystallize. We note that they are similar to the values obtained by Singh, 
Ediger, and de Pablo \cite{E_NM}. 

\begin{figure}
\includegraphics[scale=0.3]{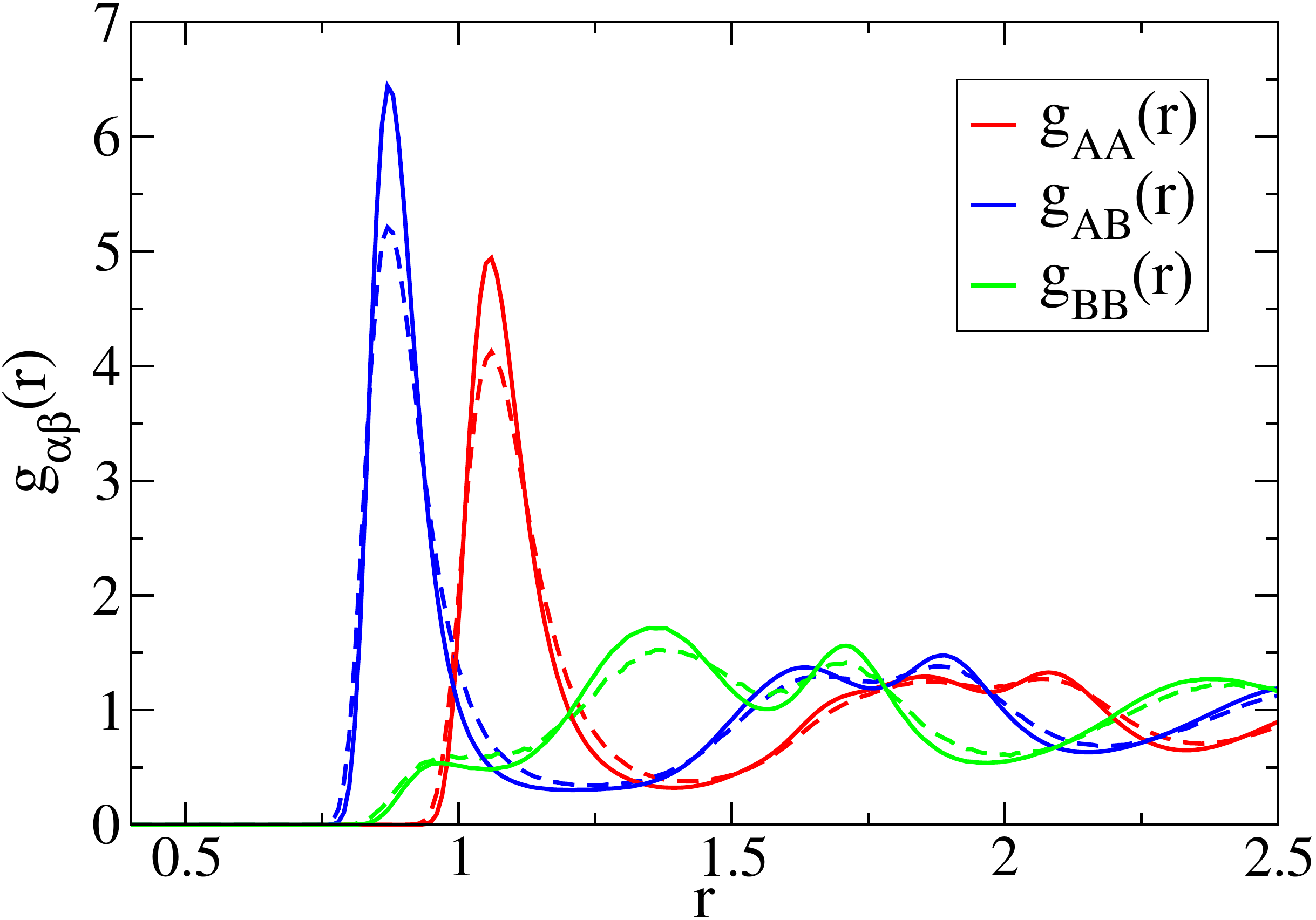}
\caption{\label{fig:gr} Solid lines: 
partial pair distribution functions measured at a temperature of $T = 0.3$ for the cooling rate of 
$\dot{T} = 3.33 \times 10^{-7}$. The pair distribution functions for the other cooling rates are nearly identical.
Dashed lines: partial pair distribution functions in equilibrium at a temperature of $T = 0.5$}
\end{figure}

As another check for crystallization we examined the partial pair distribution functions
\begin{equation}
\label{eq:gr}
g_{\alpha \beta}(r) = \frac{V}{N_{\alpha} N_{\beta}} \left< \sum_{i}^{N_{\alpha}} \sum_{j \ne i}^{N_{\beta}} 
\delta [r - (\mathbf{r}_j - \mathbf{r}_i)] \right>,
\end{equation}
where $V$ is volume and $N_\alpha$ is number of particles of type $\alpha$. 
Shown in Fig.~\ref{fig:gr} are the partial pair distribution functions $g_{AA}(r)$, $g_{AB}(r)$, and $g_{BB}(r)$ at $T=0.3$ 
after cooling (solid lines) and at $T=0.5$ in equilibrium (dashed lines). The distribution functions at $T=0.3$ are 
qualitatively similar to those of the supercooled liquid at $T = 0.5$. 
The peaks have the same locations, but are slightly more pronounced at $T = 0.3$ than 
$T = 0.5$. For both temperatures there are no indications in $g_{\alpha \beta}(r)$ that they system has crystallized 
or that the two species are no longer homogeneously distributed. 

\begin{table*}
\caption{\label{table1} The $Q_6$ parameter.}
\begin{center}
  \begin{tabular}{c @{\hspace{6mm}} c @{\hspace{6mm}} c @{\hspace{6mm}} c @{\hspace{6mm}} c } \hline\hline\noalign{\smallskip}
  Cooling rate         & $Q_6 {}$ at constant volume & standard deviation & $Q_6 {}$ at constant pressure & standard deviation \\ \hline \noalign{\smallskip}
  $3.33 \times 10^{-3}$ & 0.0257   & 0.00019            & NA       & NA                \\
  $3.33 \times 10^{-4}$ & 0.0257   & 0.00021            & 0.0229   & 0.00017           \\ 
  $3.33 \times 10^{-5}$ & 0.0258   & 0.00020            & 0.0228   & 0.00020           \\ 
  $3.33 \times 10^{-6}$ & 0.0259   & 0.00021            & 0.0227   & 0.00019           \\ 
  $3.33 \times 10^{-7}$ & 0.0261   & 0.00030            & 0.0228   & 0.00095           \\ 
  $3.33 \times 10^{-8}$ & 0.0263   & 0.00009            & 0.0229   & 0.00038           \\ \hline\hline
  \end{tabular}
\end{center}
\end{table*}

We note that one cooling run at a cooling rate of $3.33 \times 10^{-8}$ resulted in a slightly different final configuration. 
We saw no signs of crystallization in $Q_6$ or $g_{\alpha \beta}(r)$. However, when we ran the heating trajectories, 
we noticed that the potential energy initially rose towards the equilibrium value at $T = 0.5$, but then dropped. 
When we continued one of these heating trajectories (continuing the simulation at $T = 0.5$), we found, by examining partial
pair distribution functions, that the 
$A$ and $B$ particles had begun to separate. We did not use this cooling trajectory or its subsequent 
heating trajectories in our results. We note that this separation of particles suggests that we may have reached the limit of 
how slowly we can cool this system at the pressure of 3.958 and still maintain a homogeneous liquid structure.

\section{Kinetic Stability}
\label{sec:stability}
In this section we examine the kinetic stability of the glasses created by constant pressure cooling. 
To this end we heat 
these glasses to the supercooled liquid temperature of 
$T = 0.5$ at constant pressure and monitor the particles' dynamics. 
The time it takes for the system to return to the liquid state is a measure of the 
stability of the glass. We study the stability for our constant pressure simulations and compare these results to the constant 
volume simulations of Ref.~\cite{cool}. 

\subsection{Heating trajectory dynamics}
 
We examine dynamics during the heating trajectories by calculating the average overlap function,
\begin{equation}
\label{eq:qs}
q_s (t,t_w) = \frac{1}{N} \left< \sum_m q_m (t,t_w) \right>,
\end{equation}
where
\begin{equation}
\label{eq:qm}
q_m (t,t_w) = \Theta (a - \left| \mathbf{r}_m (t + t_w) - \mathbf{r} (t_w) \right|),
\end{equation}
$\Theta$ is the Heaviside step function, and $\mathbf{r}_m (t)$ is the position of particle $m$ at time $t$. The function $q_s (t, t_w)$ measures the fraction of particles that moved less than a distance $a$ from $t_w$ to $t_w + t$. The waiting time $t_w$ is measured from the beginning of the trajectory. Recall that for a system in equilibrium, the average overlap function
does not depend on waiting time $t_w$. As in previous work \cite{cool}, we use a value of $a = 0.25$. 

\begin{figure}
\includegraphics[scale=0.3]{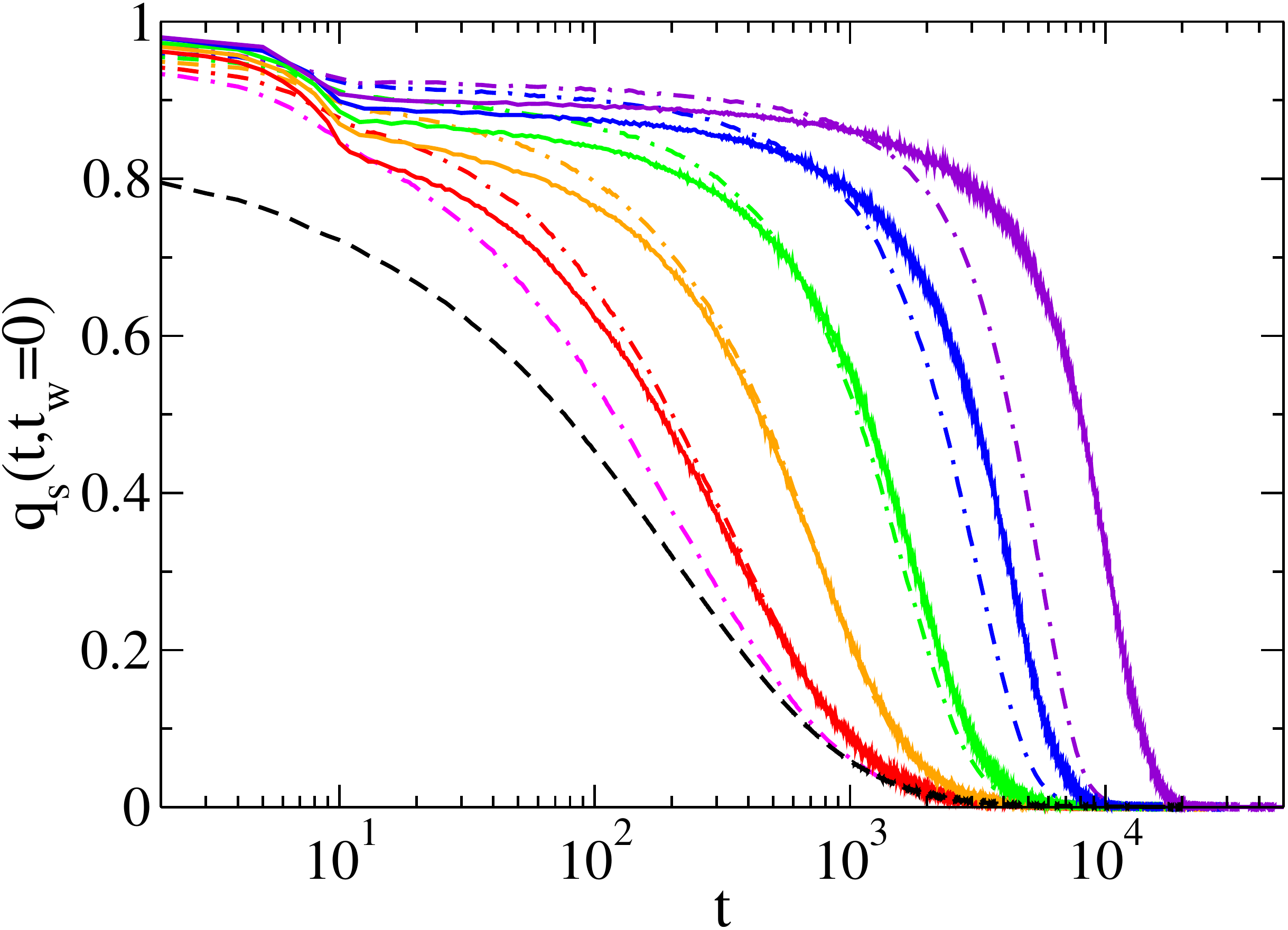}
\caption{\label{fig:theta0} The average overlap function at a waiting time of 0 for the constant pressure heating simulations (solid lines), the constant volume heating simulations (dot dashed lines), and the equilibrium simulation (dashed line) at $T = 0.5$. The cooling rates are $3.33 \times 10^{-n}$, where $n$ goes from 3 to 8 for the constant volume simulations, and from 4 to 8 for constant pressure simulations. $n$ increases from left to right. Matching values of $n$ have the same color.}
\end{figure}

Shown in Fig.~\ref{fig:theta0} is $q_s (t,t_w = 0)$ during the constant pressure heating trajectories (solid lines) for the different cooling rates. Also shown is $q_s(t,t_w)$ for the equilibrium system at $T = 0.5$ (dashed line), which is independent of $t_w$, and results from the constant volume simulations (dot-dashed lines). We note that the kink in the curves at $t = 10$ is due to the change from heating at a constant rate to holding the temperature constant. As in the constant volume simulations, there is a plateau in $q_s(t,t_w=0)$ for the smaller cooling rates, indicating that particles are trapped in cages formed by their neighbors. This plateau lengthens and its height is increasing with decreasing cooling rate. The height of the plateau is slightly lower in the constant pressure simulations than in the constant volume simulations, which suggests that the cages are slightly larger in the constant pressure simulations. This is a bit surprising since the density during the constant pressure heating runs is higher than during the constant volume heating runs. At the slowest cooling rates the plateau persists for a longer time in the constant pressure simulations than in the constant volume simulations.

\begin{figure}
\includegraphics[scale=0.3]{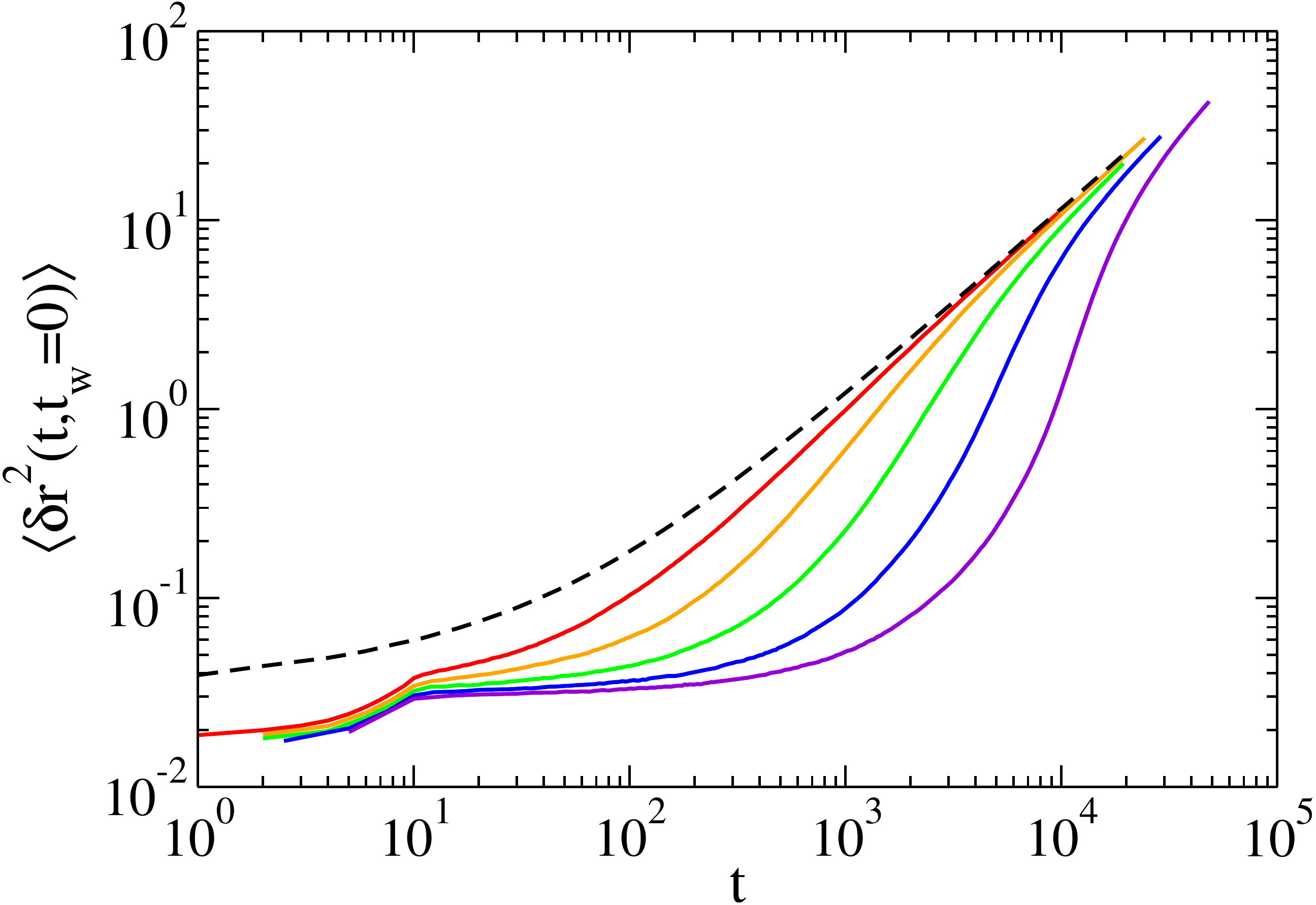}
\caption{\label{fig:msd} The mean square displacement at a waiting time of 0 (solid lines). The mean square displacement for an equilibrium fluid at $T = 0.5$ is shown with a dashed line. The cooling rates are $3.33 \times 10^{-4}$, $3.33 \times 10^{-5}$, $3.33 \times 10^{-6}$, $3.33 \times 10^{-7}$, $3.33 \times 10^{-8}$, from top to bottom.}
\end{figure}

Figure~\ref{fig:msd} shows the mean square displacement
\begin{equation}
\label{eq:msd}
\left< \delta r^2 (t,t_w) \right> = \frac{1}{N} \left< \sum_n [\mathbf{r}(t + t_w) - \mathbf{r}(t_w)]^2 \right>,
\end{equation}
for $t_w = 0$ for the constant pressure heating trajectories (solid lines), and for the $T = 0.5$ equilibrium run (dashed line). 
The features in $\left<\delta r^2(t,t_w=0) \right>$ mirror 
those in $q_s(t,t_w=0)$ shown in Fig.~\ref{fig:theta0}. 
There is a kink at $t = 10$ due to the change from heating to holding the temperature constant, and at long times, 
the heating trajectory curves begin to approach the equilibrium curve. 
The slower cooling rate glasses take longer to return to the equilibrium curve than the glasses created at faster cooling rates. 

\subsection{Stability ratio}

We quantify the kinetic stability of our glasses by defining a stability ratio $S$,
which is a measure of how long it takes the glass to return to equilibrium upon having been heated to a liquid-like temperature 
relative to the equilibrium relaxation time $\tau_{\alpha}$ at this temperature. We obtained the transformation time 
$t_{\mathrm{trans}}$ following a procedure from Ref.~\cite{HBR} and our previous work \cite{cool}. 
For this procedure we define a waiting time dependent relaxation time $\tau_s(t_w)$ through $q_s (\tau_s,t_w) = e^{-1}$. 
The transformation time $t_{\mathrm{trans}}$ is defined as the minimum $t_w$ where $\tau_s (t_w) = \tau_{\alpha}$
where $\tau_\alpha$ is the equilibrium relaxation time. 
We then define the stability ratio as $S = t_{\mathrm{trans}}/\tau_{\alpha}$. We note that this stability ratio depends on 
the temperature to which the glass is heated \cite{cool}, and we study the stability ratio for temperature $T=0.5$ in order to 
compare with our previous work \cite{cool}. 

\begin{figure*}
\includegraphics[scale=0.6]{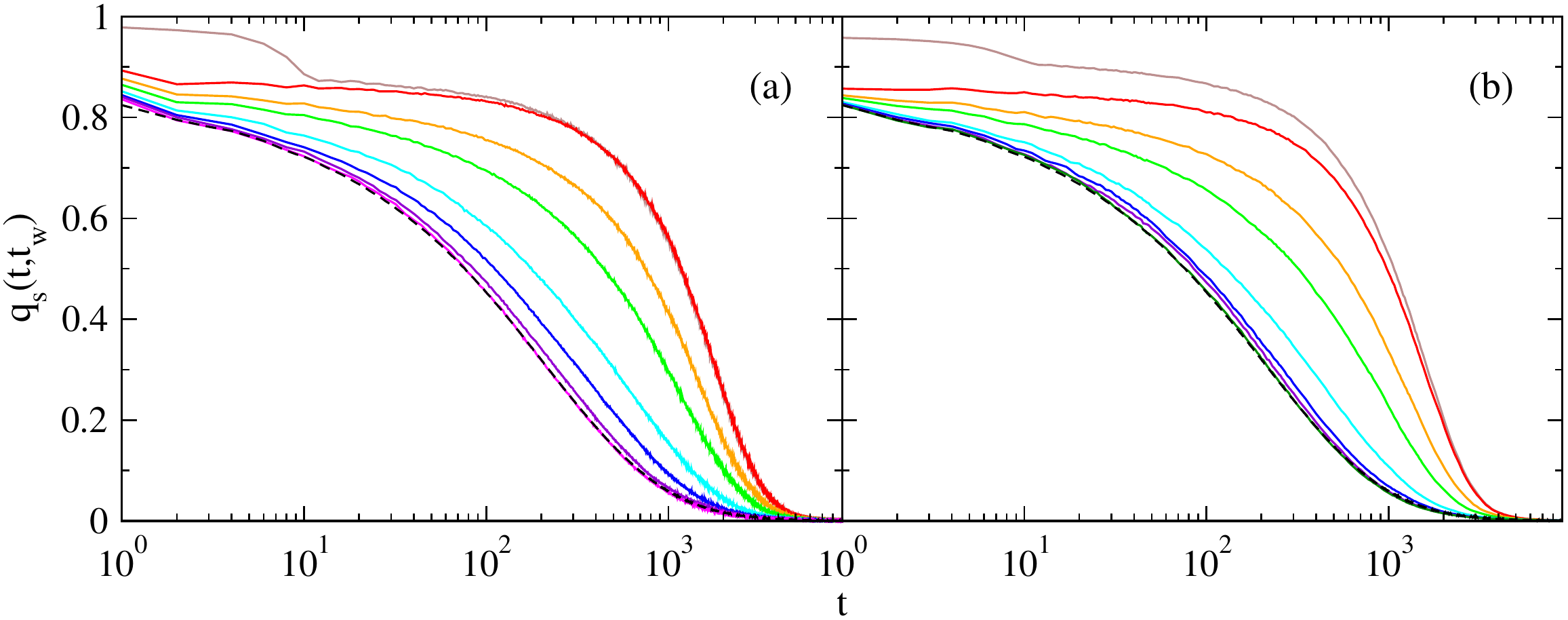}
\caption{\label{fig:thetac6} The average overlap function for the cooling rate of $3.33 \times 10^{-6}$. The left panel gives results from constant pressure simulations, and the right panel has results from constant volume simulations. The dashed lines are the average overlap function for equilibrium at $T = 0.5$. The waiting times are 0, 10, 500, 1000, 2000, 3000, 4000, 6500 from top to bottom on the left. On the right, the waiting times are 0, 10, 500, 1000, 2000, 3000, 4000, 4250 from top to bottom. Note that corresponding colors in each panel have the same waiting time and the largest waiting time in (a) is different 
than the largest waiting time in (b).}
\end{figure*}

In Fig.~\ref{fig:thetac6} we show $q_s (t,t_w)$ for heating trajectories for glasses created at a cooling rate of 
$3.33 \times 10^{-6}$. Shown in Fig.~\ref{fig:thetac6}(a) are results for the constant pressure simulations, 
and shown in Fig.~\ref{fig:thetac6}(b) are results for the constant volume simulations from Ref.~\cite{cool}. 
As the waiting time increases the curves approach the equilibrium curve for $T = 0.5$. Matching colors on each side correspond to the same waiting times. We can see from this figure that the return to equilibrium takes longer in the constant pressure simulations for this cooling rate and heating procedure.

\begin{figure}
\includegraphics[scale=0.3]{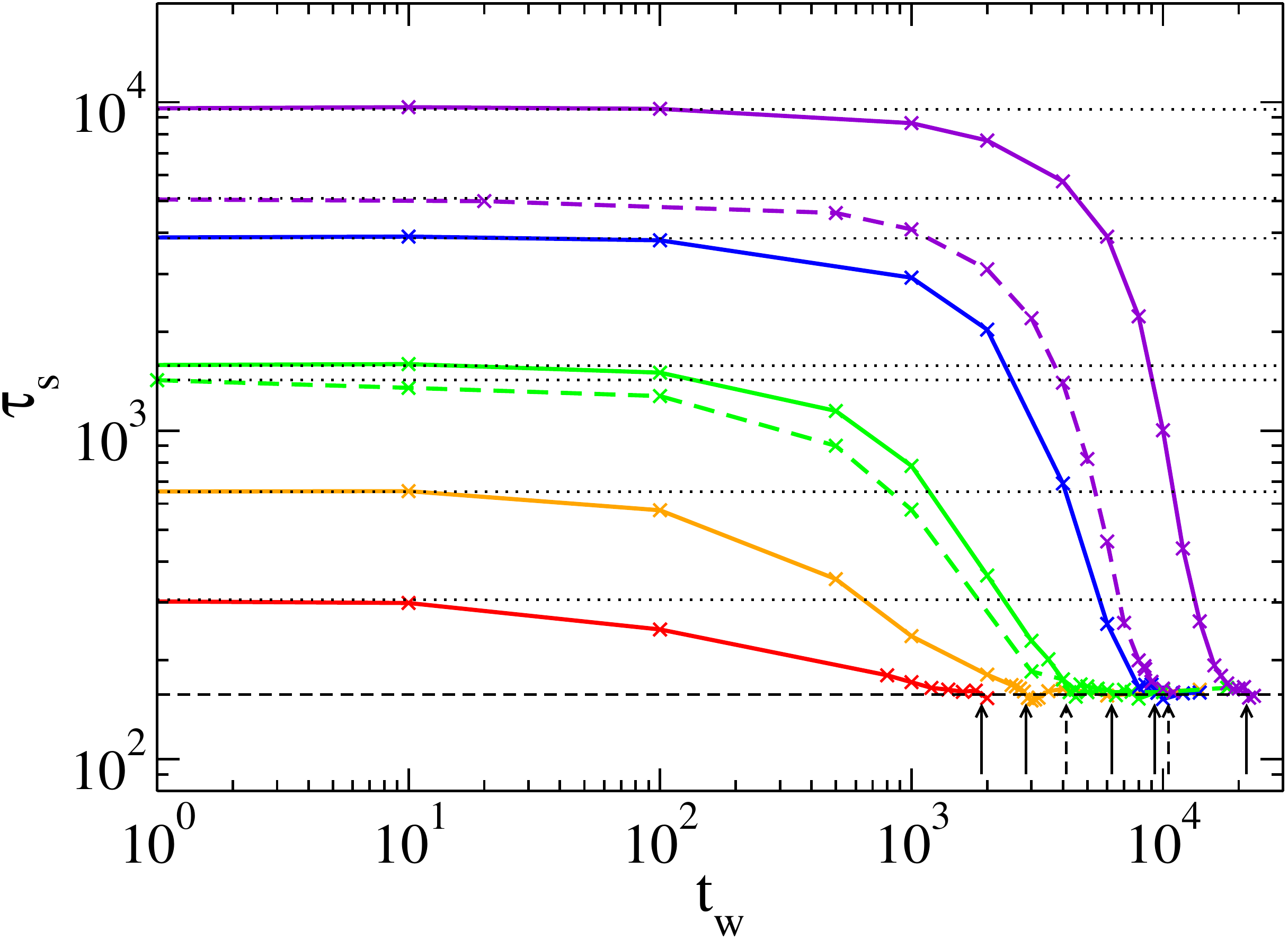}
\caption{\label{fig:ts} The out of equilibrium relaxation time plotted as a function of waiting time for the five cooling rates. Solid lines are constant pressure runs, with cooling rate decreasing from bottom to top. Dashed lines of the same color are from constant volume simulations at the same cooling rate. The black dashed line gives the equilibrium relaxation time at $T = 0.5$. The dotted lines give the initial value of $\tau_s$ at $t_w = 0$. The arrows point to the transformation times $t_{trams}$. Dashed arrows are for constant volume simulations and solid arrows are for constant pressure simulations. For the constant pressure simulations the cooling rates are $3.33 \times 10^{-n}$, where $n$ goes from 4 to 8 from bottom to top. The cooling rates of $3.33 \times 10^{-6}$ and $3.33 \times 10^{-8}$ are shown for the constant volume simulations.}
\end{figure}

Shown in Fig.~\ref{fig:ts} is the waiting time dependent relaxation time $\tau_s$ as a function of waiting time $t_w$. As can be inferred in in Fig.~\ref{fig:thetac6}, $\tau_s$ approaches $\tau_\alpha$ with increasing waiting time. Also shown is a comparison with the constant volume simulations at the cooling rates of $3.33 \times 10^{-6}$ and $3.33 \times 10^{-8}$ (dashed lines). 
The transformation time is defined as the smallest waiting time when $\tau_s = \tau_{\alpha} = 160$, and this time is marked by the arrows in the figure.

\begin{figure}
\includegraphics[scale=0.3]{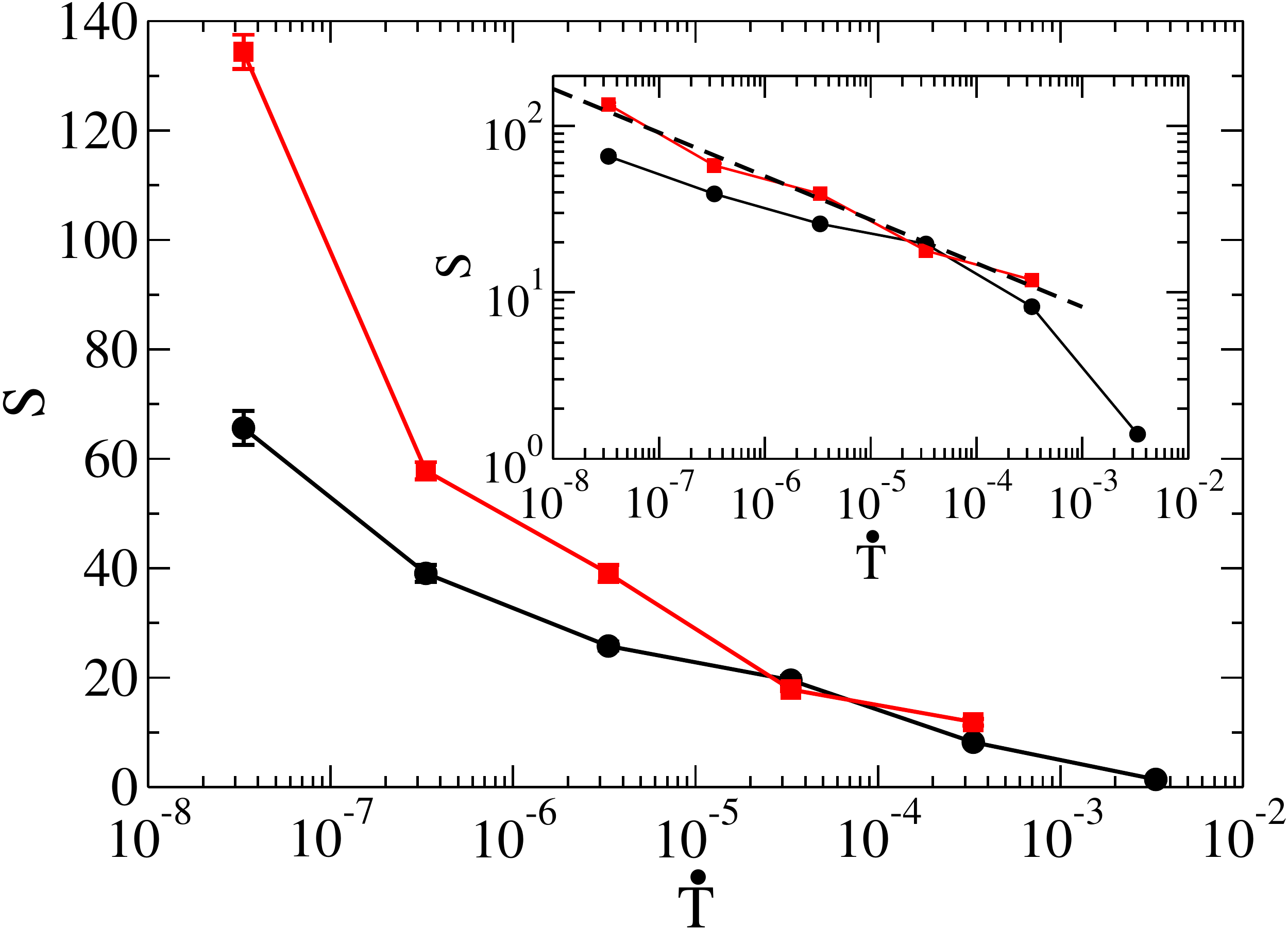}
\caption{\label{fig:S} The stability ratio $S$ versus the 
cooling rate $\dot{T}$ calculated from constant pressure cooling and heating (red squares), and 
constant volume cooling and heating (black circles). 
The inset shows the stability ratios on a log scale. 
The dashed line is a fit to $\log_{10}(S) = m \log_{10}(\dot{T}) + b$ 
where $m = -0.26$ and $b = 0.13$.}
\end{figure}

Fig.~\ref{fig:S} shows the stability ratio $S = t_{\mathrm{trans}}/\tau_\alpha$ for $T = 0.5$ as a function of cooling rate. 
The red squares are constant pressure results and the black circles are constant density results. At our two fastest cooling rates, $3.33 \times 10^{-4}$ and $3.33 \times 10^{-5}$, the stability ratios are nearly identical. 
However, at the three slowest cooling rates the constant pressure stability ratios are larger than 
the constant volume stability ratios, and the constant pressure stability ratio increases
faster with decreasing cooling rate than the constant volume stability ratio.

We fit the stability ratio to $\log_{10}(S) = a\log_{10}(\dot{T}) + b$, 
and obtained $a = -0.26$ and $b = 0.13$. 
Extrapolating the fit to the stability ratio of the most stable simulated glasses, $S \simeq 400$ \cite{HBR}, we find
that one would have to cool the system 2 orders of magnitude slower than our slowest cooling rate ($3.33 \times 10^{-8}$) 
to match this stability ratio. 
Extrapolating the fit to the stability ratio 
of experimental ultrastable glasses created by vapor deposition, which have an stability ratio of 
$S \simeq 10^{3.5}$ \cite{Sep}, we would need to cool our glasses 5 orders of magnitude slower than our slowest cooling rate. 
As we noted in Subsection~\ref{sec:crystal}, we appear to be at the limit of our cooling rate without fractionation 
and/or crystallization intervening.

We note that compared to our previous study \cite{cool} we not only changed the simulation method to create the glass, but
also the method used to melt the glass. To determine if the reheating procedure 
changes the stability ratio we reheated the glass created by cooling 
at $\dot{T} = 3.33\times10^{-7}$ at constant pressure using two alternative procedures. We heated the glass at constant 
volume at density $\rho = 1.2451$ (which was the density at the end of the constant pressure cooling runs) 
to $T=0.5$ and to $T=0.6$. We choose $T=0.6$ since 
$\tau_\alpha$ at $T=0.5$ for the commonly used density $\rho = 1.2040$ is nearly equal to $\tau_\alpha$ 
at $T=0.6$ for $\rho = 1.2451$. We found that the constant volume reheating resulted
in a smaller stability ratio ($S \approx 2$) when reheating to $T=0.5$.
However, the stability ratio for the reheating to 
$T=0.6$ while maintaining density at $\rho = 1.2451$ was slightly larger ($S \approx 50$) than the constant volume
stability ratio for $\rho = 1.2040$ when reheating to $T=0.5$, but it is smaller than the 
constant pressure stability ratio.  We recall that we previously found that the stability ratio 
depended on the temperature to which we reheated \cite{cool}.
Further work would be needed to understand how the 
stability ratio is related to the reheating procedure. However, it is clear
that comparisons of stability ratios using different procedures to create and 
reheat simulated glasses should be done with care. 

\begin{figure}
\includegraphics[scale=0.3]{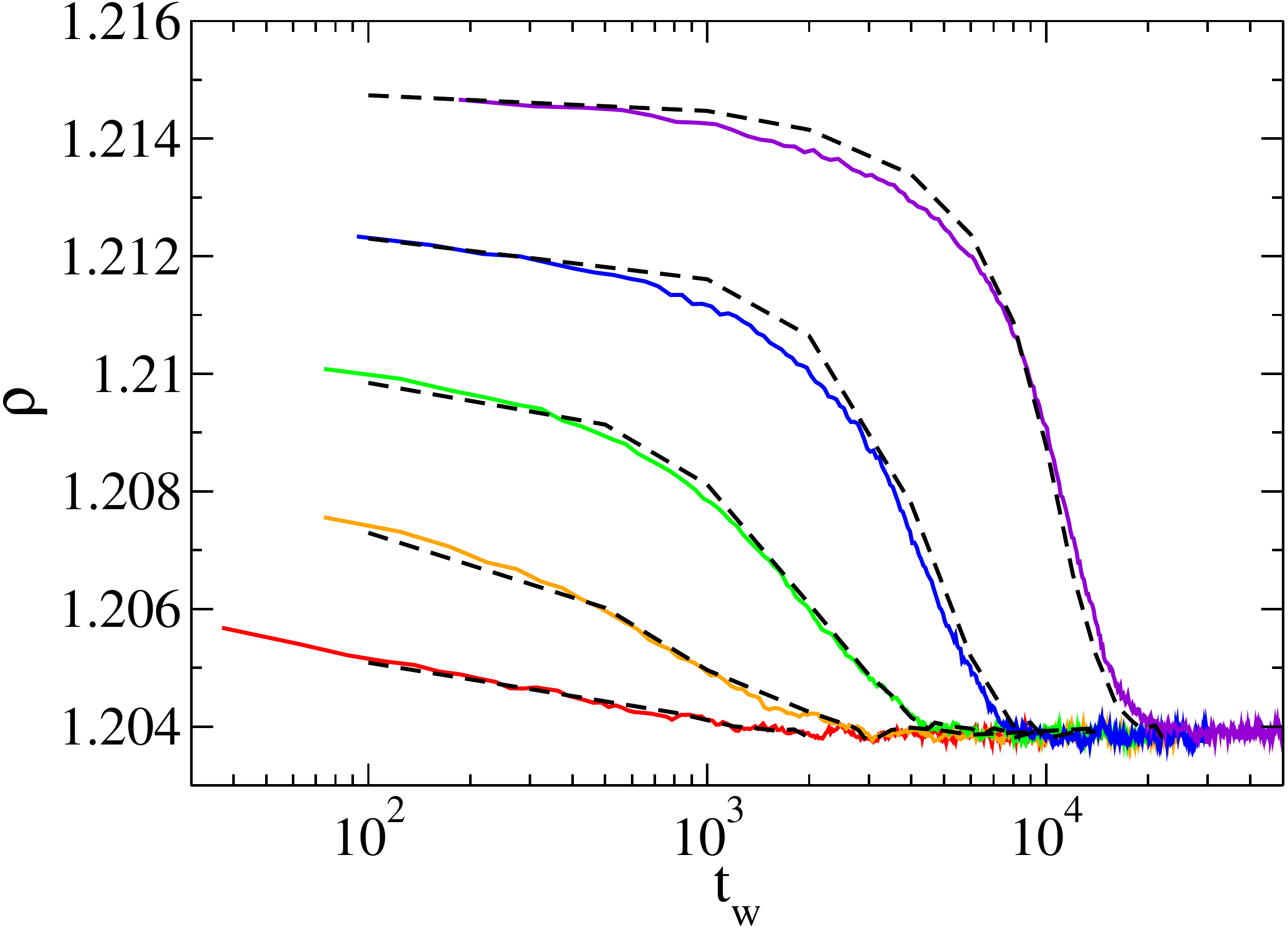}
\caption{\label{fig:dens_ts} Density $\rho$ (solid lines) and rescaled out of equilibrium relaxation time $\tau_s$ (black dashed lines) plotted as a function of waiting time $t_w$. The out of equilibrium relaxation times were scaled by $0.00609 \log_{10} (\tau_s) + 1.1905$. The different color curves represent different initial cooling rates of $\dot{T} = 3.33 \times 10^{-n}$, where $n$ goes from 4 to 8 from bottom to top. Note that the scaling is independent of cooling rate.}
\end{figure}

For the constant pressure simulations the box volume changes in response to the the pressure, and this volume change results in a change in the density. We also examined how the density changed for the different cooling rates as the glass transformed back into a supercooled liquid. To examine how the density change 
is related to $\tau_s$ we compared $\rho = N/V$ to $\tau_s$ in Fig.~\ref{fig:dens_ts}. 
To facilitate this comparison we rescaled $\tau_s$ using 
$0.00609 \log_{10} (\tau_s) + 1.1905$ (solid lines) and compared
these rescaled $\tau_s$ as a function of waiting time $t_w$ to $\rho$ as a function of $t_w$. The rescaled out of equilibrium relaxation times curves closely match the density curves, suggesting that the return to equilibrium dynamics is correlated with a change in the density. Furthermore, this provides an easier method to determine the transformation time in constant pressure simulations; all one has to monitor is the density as a function of 
time instead of calculating $q_s(t,t_w)$ for many different waiting times. 

\section{Energy and density of the glass}\label{sec:energetics}

Simulations have shown that as a liquid is supercooled it spends more time around lower 
inherent structure energy minima \cite{Berthier2011}.  
Simulations have also provided evidence that the stability 
of a glass is related to the average inherent structure energy \cite{JHGC,Helfferich2016,HBR,SdP,E_NM,E_JCP,cool}, with 
glasses with a lower average inherent structure energy being more stable.   
In this section we examine the average potential energy and the 
average inherent structure energy for the glasses at $T = 0.3$. We compare the results for glasses obtained by 
the constant pressure and constant volume cooling as a function of the cooling rate. We find that the average potential energy
and inherent structure energy are lower for the constant pressure simulations at a 
given cooling rate. Moreover, we also find that the 
average potential energy and inherent structure energy are linearly related and this relationship is statistically
independent of whether the glass is cooled at constant pressure or constant volume, even though the 
density increases for the constant pressure simulations.   

\begin{figure}
\includegraphics[scale=0.6]{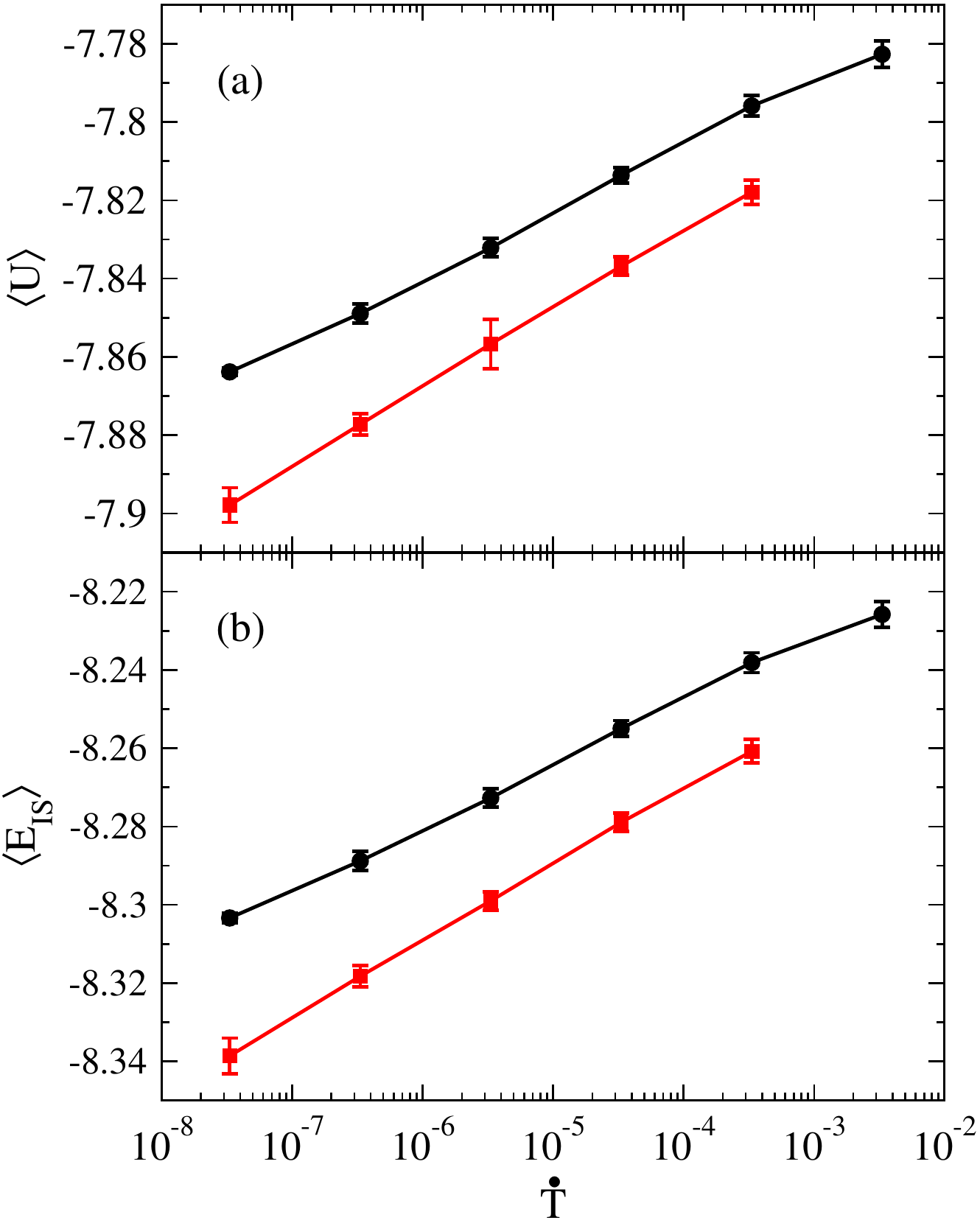}
\caption{\label{fig:UEISpanels} (a) The potential energy at $T = 0.3$ 
for constant volume cooling (black circles) and constant pressure cooling (red squares). 
(b) The inherent structure energy at $T=0.3$ for constant volume cooling (black circles) and constant pressure cooling (red squares).}
\end{figure}

Figure~\ref{fig:UEISpanels}(a) shows the average potential energy $\left< U \right>$ and
Fig.~\ref{fig:UEISpanels}(b) shows the average inherent structure energy $\left< E_{IS} \right>$ of the glasses cooled at 
constant pressure (red squares) and constant volume (black circles). 
For both constant volume and constant pressure, $\left< U \right>$ and $\left< E_{IS} \right>$
decreases as cooling rate decreases,
which suggests that the lower average energy indicates a more stable glass. 
The constant pressure results are lower than the constant volume results at each cooling rate. Note that, however, 
the stability ratio for the $\dot{T} = 3.33\times10^{-5}$ and $\dot{T} = 3.33\times10^{-4}$ are 
nearly the same for constant pressure and constant volume. Therefore, $\left< U \right>$ 
and $\left< E_{IS} \right>$ 
should not be used solely as a measure of stability, and they are only suggestive of a more stable glass. 

Helfferich \textit{et al.}\ \cite{Helfferich2016} found that the inherent structure energy was a good indicator of the mobility of 
particles in a glass film, and that an aged film with the same average inherent structure energy as a vapor 
deposited film has the same dynamics. Helfferich's results suggest the the inherent structure energy could 
be used as a measure of the stability of the system, but, taken together with our results, we find that the inherent
structure energy is not a sole measure of the stability.
However, at a fixed pressure or fixed volume, the inherent structure energy is correlated with the stability. 

\begin{figure}
\includegraphics[scale=0.3]{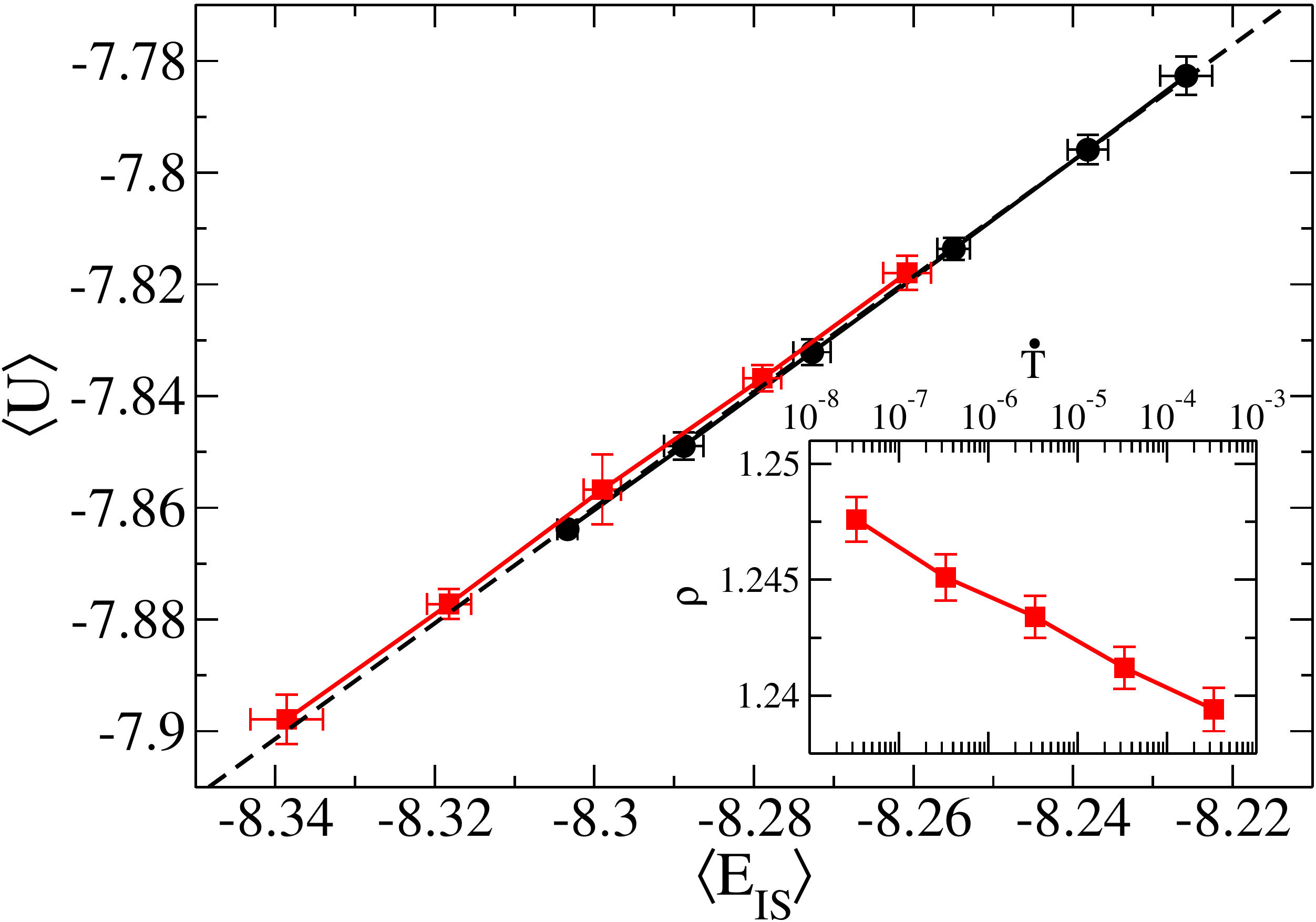}
\caption{\label{fig:UErho}The average potential energy $\left< U \right>$ versus the 
average inherent structure energy $\left< E_{IS} \right>$ at 
$T=0.3$ for cooling at constant volume (black circles) and constant pressure (red squares). 
The dashed line is a linear fit
$\left< U \right> = m \left< E_{IS} \right> + U_0$ where $m = 1.035 \pm 0.001$ and $U_0 = 0.73 \pm 0.09$.
The inset shows the density at $T=0.3$ as a function of cooling rate for the constant pressure simulations. Despite 
an increase in the density, the relationship between $\left< U \right>$ and $\left< E_{IS} \right>$ does not
change.}
\end{figure}

The similarity between the cooling rate dependence of $\left< U \right>$ and $\left< E_{IS} \right>$ motivated 
us to examine the relationship between $\left< U \right>$ and $\left< E_{IS} \right>$. Shown in Fig.~\ref{fig:UErho} 
is $\left< U \right>$ versus $\left< E_{IS} \right>$ for the constant volume simulations (black circles) and 
the constant pressure simulations (red squares). Despite the change in density for the constant pressure 
simulations (see the inset to Fig.~\ref{fig:UErho}), the relationship between $\left< U \right>$ and $\left< E_{IS} \right>$ 
remain unchanged to within the statistical uncertainty of our results. Furthermore, a fit to 
$\left< U \right> = m \left< E_{IS} \right> + U_0$ results in $m = 1.035 \pm 0.001$ and $U_0 = 0.73 \pm 0.09$. Note
that the relationship between the two is linear with a slope nearly equal to one and independent of a changing density 
for the constant pressure simulations. 

Lyubimov \textit{et al.} \cite{E_JCP} modeled the vapor deposition process using the Kob-Andersen system studied in the 
present investigation. They used a substrate temperature of 0.3, and found that their vapor deposited simulated glasses 
had an average potential energy $\left< U \right>$ of -7.8 and an average inherent structure energy $\left< E_{IS} \right>$ 
of -8.35. While this does not fit our relationship between potential and inherent structure energies, 
we note that in the study of Ref. \cite{E_JCP} the density of the vapor deposited film changes during the energy minimization. 
Thus, the density of the system used to calculate the average potential energy is different from the density of the system
used to calculate the inherent structure energy.  
The present procedure of cooling at a constant pressure has resulted in the average inherent structure energy 
at the slowest cooling rate being 
much closer to the value of Lyubimov \textit{et al.} than in our constant volume simulations. We note that, however, 
the lower inherent structure energy does not necessarily indicate an increase in the stability ratio. 

\section{Summary and Conclusions}
\label{sec:conclusions}

We examined the change in the stability ratio  $S = t_{\mathrm{trans}}/\tau_\alpha$
when we changed the method to create and melt 
a model glass forming system. We find that for a slow enough cooling rate, that the stability ratio
of a system cooled and heated under constant pressure conditions is larger than the stability
ratio of a system cooled and heated under constant volume conditions.
We found that we would still need to cool at a rate two 
orders of magnitude slower to reach the most stable glass formed in simulations,
and five orders of magnitude slower to equal the stability of glasses
formed in the laboratory by vapor deposition. We also found that the stability ratio was sensitive not 
only to the cooling procedure used to create the glass, but the heating procedure used to melt the
glass. Therefore, care must be taken in comparing stability ratios obtained by different methods. 

One unexpected feature of melting the glass at constant pressure was that the plateau height of the 
average overlap function was lower for constant pressure, but the relaxation time was longer. 
These results suggest that the particles have more room to move, despite having a smaller
average volume, but the glass takes longer to melt.  

While studying the melting of the glass under constant pressure, we found that the time dependence
of the density was a good indicator of the transition time. Monitoring the volume change is a more 
efficient method to determine the transition time than finding the waiting time when the overlap function 
has the same decay time as for the equilibrium bulk sample. Future work on simulated glass films 
should also examine the density change upon melting and examine if 
there is a heterogeneous density change 
starting at the surface, and if this density change is related to the melting
of stable glasses due to a mobile front initiated at the surface \cite{Sep,Swallen2009,Swallen2011}. 

We also examined the average potential energy $\left< U \right>$ and the average inherent structure 
energy $\left< E_{IS} \right>$ for the glasses. Both of these quantities are frequently used as indicators of the 
stability of the glass \cite{JHGC,Helfferich2016,HBR,SdP,E_NM,E_JCP,cool}, 
and it has been shown that the particles mobility in a glass forming film 
is correlated with $\left<E_{IS}\right>$ \cite{Helfferich2016}.  
Since the two fastest cooling rates for the systems prepared at constant volume and constant 
pressure had the same stability ratio but different $\left< U \right>$ and $\left< E_{IS} \right>$, we conclude 
that the stability cannot be inferred from these quantities alone. However, $\left< E_{IS} \right>$ is correlated
with the stability ratio if either the glass is prepared and melted and constant volume or constant pressure. To 
understand the stability of glasses created by simulated vapor deposition, one would  then need to examine
the stability of simulated glass films created through vapor deposition, films created at constant zero pressure,
and bulk simulations at constant zero pressure. 
It is still unclear how the free surface and the substrate are going to influence the calculation of a stability ratio 
for simulated vapor deposited films. 

We gratefully acknowledge the support of NSF grant CHE 1213401.

\end{document}